\documentclass[conference]{IEEEtran}
\IEEEoverridecommandlockouts
\usepackage{cite}
\usepackage{amsmath,amssymb,amsfonts}
\usepackage{algorithmic}
\usepackage{graphicx}
\usepackage{textcomp}
\usepackage{xcolor}
\def\BibTeX{{\rm B\kern-.05em{\sc i\kern-.025em b}\kern-.08em
    T\kern-.1667em\lower.7ex\hbox{E}\kern-.125emX}}
\begin{document}

\title{Open5G: A Software-Defined Networking Protocol for 5G Multi-RAT Wireless Networks}

\author{
	\IEEEauthorblockN{Pradnya Kiri Taksande\IEEEauthorrefmark{1},
		Pranav Jha\IEEEauthorrefmark{1},
		Abhay Karandikar\IEEEauthorrefmark{1}\IEEEauthorrefmark{2},
		Prasanna Chaporkar\IEEEauthorrefmark{1},
	}
	\IEEEauthorblockA{\IEEEauthorrefmark{1}Department of Electrical Engineering,
		Indian Institute Technology Bombay, India 400076\\
		Email: \{pragnyakiri,pranavjha,karandi,chaporkar\}@ee.iitb.ac.in}
	\IEEEauthorblockA{\IEEEauthorrefmark{2}Director, Indian Institute Technology Kanpur, India 208016\\
		Email: karandi@iitk.ac.in
	}
}

\maketitle

\begin{abstract}
Mobile Networks today comprise of multiple Radio Access Technologies (RATs), e.g., 4G Long Term Evolution (LTE), Wireless Local Area Network (WLAN), and the upcoming 5G-New Radio (5G-NR). The access networks of these RATs are controlled by RAT-specific entities, e.g., the resource management function located inside an individual LTE eNB is used for the eNB control, or access controllers are used for controlling WLAN Access Points. Even in the 3rd Generation Partnership Project's (3GPP) 5G architecture, which has a common Core supporting multiple RATs, radio access related decisions are taken independently within individual RATs. Due to the fragmented nature of control-plane in multi-RAT Radio Access Network (RAN), a unified global view of network resources is unavailable, hindering optimized allocation of resources. It also brings complexity to the features involving multiple RATs, e.g., dual connectivity. To address these issues, we introduced an SDN-based Multi-RAT RAN architecture (SMRAN) in our earlier work \cite{Pradnya}, where the RAN control-plane is segregated from the data-plane. As part of the SMRAN architecture, we defined a logically centralized multi-RAT RAN Controller and individual RAT-specific data-plane functions. In the current work, we define a protocol, called Open5G, to be used for control and management of the SMRAN data-plane. Open5G is based on OpenFlow (OF) and OF-Config, which are commonly used protocols in the SDN-based wired networks and data centers. With the Open5G protocol, the multi-RAT RAN can be controlled by an open interface, bringing flexibility and simplicity in network interactions. 
\end{abstract}


\section{Introduction}
With improving connectivity and accessibility to smart devices, the demands for mobile data traffic are rising day by day. One of the approaches to meet these demands is to deploy small cells alongside a homogeneous macro cell network giving rise to heterogeneous networks. In a heterogeneous network, the cells may belong to multiple Radio Access Technologies (RATs). The fifth generation of cellular networks (5G) initiated by the 3rd Generation Partnership Project (3GPP) is expected to meet these increasing demands (enhanced Mobile BroadBand (eMBB)) along with providing massive Machine Type Communications (mMTC) and Ultra-Reliable Low Latency Communications (URLLC) services. Software-Defined Networking (SDN) is an emerging networking paradigm to control and manage a network effectively. It brings flexibility in the network by decoupling the control-plane from the data-plane. The concept of SDN has been applied in the 5G network, both at Core and Radio Access Network (RAN) by separating their control-plane functionality from the data-plane functionality. \par
5G RAN, known as the Next-Generation Radio Access Network (NGRAN), is made up of next-generation NodeBs (gNBs) and next-generation LTE eNodeBs (ng-eNBs). A gNB is divided into a Central Unit (gNB-CU) and a Distributed Unit (gNB-DU) consisting of centralized functions and distributed functions, respectively. gNB-CU is further subdivided into a control-plane part (gNB-CU-CP) and a user-plane part (gNB-CU-UP); thus, decoupling control-plane and data-plane functions. Communication between these different nodes of gNB takes place via multiple interfaces, viz., F1-C, F1-U, E1, etc. Each of these interfaces is defined differently by 3GPP \cite{38470,38460}. However, there is no separation of control-plane and data-plane in ng-eNBs. Moreover, even though gNB-DU is primarily a data-plane function (as defined in \cite{38401}), it is capable of taking decisions under certain scenarios. For instance, gNB-DU sends the lower layer configurations to gNB-CU when a User Equipment (UE) is admitted to the system, whereas gNB-CU sends a request to setup UE context at gNB-DU \cite{38401}. Thus, the concept of SDN is not completely applied in NGRAN.\par
Wireless Local Area Network (WLAN), which is widely deployed and used by service providers and users alike due to its ease of deployment, is integrated by 3GPP with 5G Core (5GC) via Non-3GPP Interworking Function (N3IWF). However, WLAN is not integrated within NGRAN and is typically controlled by access controllers. Therefore, nodes belonging to different RATs are not unified in NGRAN. This leads to complicated procedures for UEs with multiple RAT interfaces. UEs need to support different mechanisms compliant with different RATs in order to support interworking between multiple RATs. Consider the case of a UE dual connected to 5G (master) and LTE (secondary) nodes. If it moves out of the coverage area of its secondary node and enters the coverage area of a WLAN access point, there is no provision of dual connectivity between 5G and WLAN. Hence, the UE may lose its connection with the secondary node. Despite the usage of SDN, the current 3GPP 5G RAN architecture lacks a unified control and interworking mechanism across multiple RATs. Protocols have been defined between the control-plane and the data-plane; however, there is no standard protocol for control and management of data-plane functions belonging to different RATs. \par
Another problem with 3GPP NGRAN is the lack of logical separation between the network (data-plane) control and the UE control functionality, while separate functions for network and UE control exist in 5GC. 5GC achieves this through distinct Access and Mobility Management Function (AMF) and Session Management Function (SMF). The AMF is primarily responsible for the UE authentication and control function (e.g., through the exchange of Non-Access Stratum (NAS) messages with UE), whereas the SMF is responsible for the Network (data-plane) Control function, controlling the User Plane Function (UPF). However, there is no such separation in the NGRAN. Moreover, the F1 Application Protocol (F1AP) defined between the gNB-CU and the gNB-DU carries both the control plane messages directed towards the gNB-DU as well as the signaling messages from/to the UE. Keeping separate protocols/mechanisms to carry Network Control and UE signaling messages may facilitate a simpler architecture as these functionalities are independent of each other.\par
In this paper, we introduce an SDN-based Multi-RAT RAN (SMRAN) architecture where the control-plane of RAN nodes belonging to multiple RATs is segregated from their data-planes to address the aforementioned problems in existing 5G architecture. In our previous work \cite{Pradnya}, we defined a high-level architecture to manage and control multi-RAT RAN, and we provided some details of the functionality. However, we did not provide details of the protocol between control-plane and data-plane. In this paper, we introduce the SMRAN architecture for 5G and propose a new protocol (Open5G) for communication between the control-plane and data-plane of NGRAN. \par
In \cite{mobileflow,softair}, the authors propose an SDN-based architecture for mobile networks, where both the core and access network are modified to create an SDN-based network design. These works \cite{mobileflow,softair} propose to use OpenFlow (OF) and OF like protocol in mobile networks. However, sufficient details on the protocol and how it can be used in mobile networks have not been provided. Moreover, these works do not deal with multi-RAT RAN. O-RAN initiative \cite{oran} undertaken by the O-RAN alliance appears to be making progress in this direction. However, a unified multi-RAT RAN architecture, including 5G-NR, LTE, and WLAN is not yet available. 5G-Empower \cite{5gempower} tries to bring multiple RATs under a common platform. However, the work focuses on the interface between the Management/Application Plane and the Control Plane whereas our work is focussed on the interface between the Control plane and the Data plane. There are additional works, which define SDN based architectures for 4G or 5G RAN \cite{flexran,akshatha,hierarchical}. However, these are RAT specific proposals and do not discuss multi-RAT RAN. The paper \cite{akshatha2} defines an end-to-end SDN-based architecture for multi-RAT mobile networks. However, it proposes to use the existing 3GPP F1AP/E1AP protocols only to control the RAN data plane entities.\par
%
The novel contributions of this paper are as follows:
\begin{itemize}
	\item We introduce an enhanced version of the architecture proposed in \cite{Pradnya} as SMRAN. 
	\item In SMRAN, a logically centralized multi-RAT controller manages and controls RAN nodes belonging to different RATs.
	\item We propose Open5G as the protocol for the southbound interface between the  controller and data-plane nodes of SMRAN. It is used for the control and management of SMRAN data-plane nodes.
	\item Open5G protocol is based in OF and OF-Config protocols. To the best of our knowledge, this is the first work which explains how OF can be adapted for usage in 5G radio access networks.
	\item We also detail how the data-plane of SMRAN is configured to carry user-specific data as well as signaling. In contrast to the existing 5G architecture, SMRAN unifies the mechanism to carry UE data and control signaling.
	\item The architecture of the controller is also detailed out in this work, wherein we propose separate network and user control modules. 
	\item 
	As indicated above, Open5G is used only for the control and management of the data-plane functions. This is different from 5G F1AP protocol, where the network data-plane control and UE control messages, both are carried over F1AP.
\end{itemize}
The paper is organized as follows. In Section \ref{sec2}, we describe SMRAN. The Open5G protocol is characterized in Section \ref{sec3}. In Section \ref{sec4}, we show the working of Open5G protocol. Section \ref{sec5} describes the advantages of SMRAN and Open5G. We conclude the paper in Section \ref{conc}.\par
\section{SMRAN: SDN-based Multi-RAT RAN architecture for 5G}
\label{sec2}
%
%
\begin{figure}[!htb]
	\centering
	\includegraphics[width=8cm]{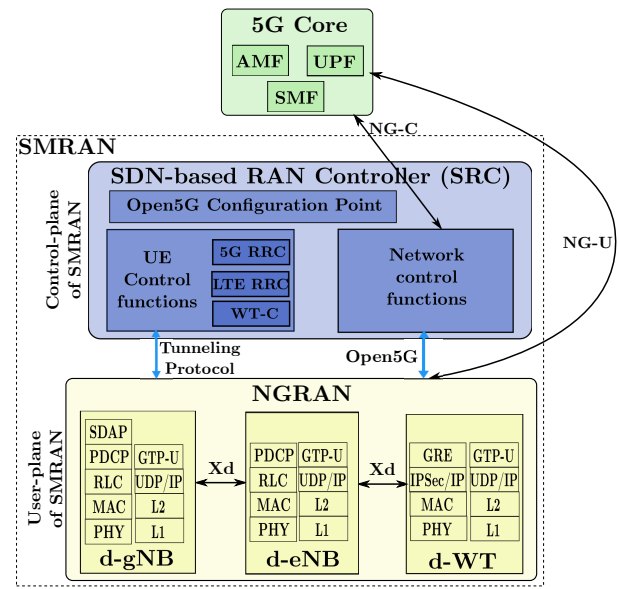}
	\caption{SMRAN: SDN based Multi-RAT RAN architecture for 5G.}
	\label{arch}
\end{figure}
In this section, we introduce SMRAN architecture for 5G, as illustrated in Figure \ref{arch}. SMRAN communicates with the 5GC through the same standard interfaces (NG-C and NG-U) as the existing 3GPP architecture, and 5GC remains unchanged. SMRAN consists of (i) NGRAN comprising data-plane entities belonging to different RATs, and (ii) an SDN-based RAN Controller (SRC) comprising the control-plane functions. The data-plane entities d-gNB, d-eNB, and d-WT are the data-plane functions of gNB, eNB, and WT, respectively. SRC incorporates an Open5G configuration point, UE control functions, and network control functions. Open5G configuration point communicates with the RAN nodes and configures them. The network control functions of SRC use the stack Next-Generation Application Protocol (NG-AP), Stream Control Transmission Protocol (SCTP), Internet Protocol (IP), Layer 2 (L2) and Layer 1 (L1) protocols to communicate with 5GC via the Next Generation Control-plane (NG-C) interface. The network control messages to configure the RAN nodes are sent using Open5G. The UE-specific control functionality of gNB (5G RRC), eNB (LTE RRC), and WT (WT-C) is hosted under UE control functions in the SRC. The control messages between the SRC and RAN data-plane nodes are carried over a tunneling protocol (GRE or it can be other tunneling protocol also) and then over LTE/5G-NR Signaling Radio Bearers (SRBs) over the radio interface. \par
The data-plane entities contain the radio protocol stack to communicate with UEs. For instance, d-gNB contains Service Data Adaptation Protocol (SDAP), Packet Data Convergence Protocol (PDCP), Radio Link Control (RLC), Medium Access Control (MAC), Physical (PHY) protocol stack, and d-WT contains MAC, PHY stack. d-WT may also contain Generic Routing Encapsulation (GRE) and IP Security protocol (IPSec) towards the UE as has been proposed in 3GPP standards for non-3GPP access interworking with 5G.
The data-plane entities also contain the GPRS Tunnelling Protocol for User Plane (GTP-U) stack to communicate with other data-plane entities and for data exchange with 5GC over the Next Generation User-plane (NG-U) interface. Xd is a common interface for communication between data-plane nodes of different RATs using GTP-U, User Datagram Protocol over Internet Protocol (UDP/IP), L2, L1 protocol stack. Open5G protocol is used by the SRC to control and manage NGRAN data-plane nodes. \par
\begin{figure}[!htb]
	\centering
	\includegraphics[height=5cm]{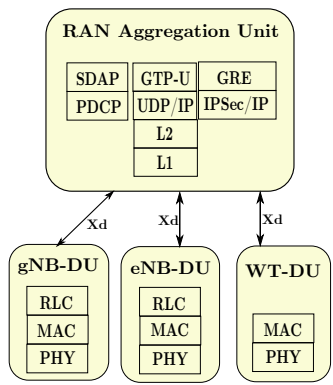}
	\caption{RAN nodes.}
	\label{RAN}
\end{figure}
In the existing 5G architecture, gNB-CU-UP and gNB-DU can be combined to obtain d-gNB in SMRAN. This alleviates the need for multiple interfaces between the different nodes. However, as illustrated in Figure \ref{RAN}, we can have a common RAN aggregation unit for aggregating the data from RAN nodes belonging to different RATs and communicating the same to the 5GC. Thus, the concept of a centralized and distributed unit in 5G can be incorporated in SMRAN as well. Thus, SMRAN is in alignment with the 5G architecture.
\section{Open5G: Open Protocol for Southbound Interface}
\label{sec3}
In this section, we describe Open5G - open protocol for the interface between SRC and NGRAN. Open5G is based on OF and OF-Config protocols with some modifications. OF \cite{OF} is an open protocol that has been introduced by the Open Networking Foundation (ONF) for communication between the control-plane and forwarding-plane in an SDN-based network \cite{OnfSdn}. OF-Config protocol \cite{OF-C} was introduced as a companion protocol to OF for configuring the forwarding-plane entities so that the controller can control the forwarding-plane entities via OF protocol. These protocols, however, are focused on wired networks and their features not directly applicable to wireless networks. There have been limited efforts \cite{mimoOF,openroads,mobileflow,softair} in adopting these protocols for wireless networks. \par
In OF, a physical port corresponds to a virtual slice of a hardware interface of the OF switch and logical port is a higher-level abstraction that is defined on top of physical ports \cite{OF}. The OF logical port is similar to the OF physical port but may have a tunnel-id as an extra meta-data field associated with it. The basic function of OF is based on flow tables, which are similar to routing tables to route the traffic from one port to another or to process certain packets. An OF switch is configured using OF-Config by an OF Configuration Point entity. We map these OF concepts to wireless networks in SMRAN, in order to define the Open5G interface. The data-plane entities can be regarded as Open5G switches, and SRC as the Open5G controller and OF Configuration Point.\par
For a UE, the signaling messages are carried from the 5GC to the UE via paths called signaling paths, and the data is exchanged between 5GC and UE by establishing PDU session paths. Since paths between SRC and 5GC are established via standard 3GPP interfaces, the main functions of Open5G protocol are 1) establishment of PDU session path between UE and radio data-plane entities and 2) establishment of signaling path between UE and SRC. Once the data and signaling paths are established, control signaling information (through signaling paths) and data (through PDU session paths) can be exchanged between the SRC/CN and UEs. In 5G, each UE has one or more PDU sessions. PDU sessions consist of data radio bearers (DRBs) from UE to gNB and a tunnel path from gNB to 5GC \cite{38300}. One PDU session can have multiple DRBs as well as multiple QoS flows. At the same time, one DRB can serve multiple QoS flows. \par
\begin{figure}
	\centering
	\includegraphics[width=7cm]{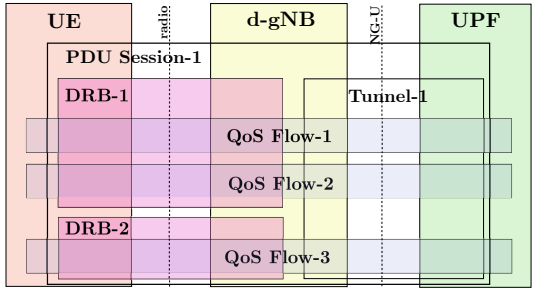}
	\caption{PDU session.}
	\label{PDUsession}
\end{figure}
\begin{figure}
	\centering
	\includegraphics[width=8.6cm]{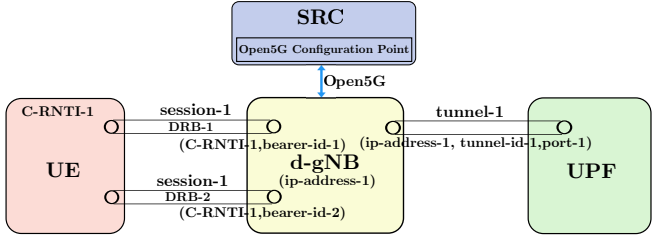}
	\caption{Open5G ports configuration for PDU session.}
	\label{mapping}
\end{figure}

Consider the establishment of a PDU session between 5GC and UE via a d-gNB, as illustrated in Figure \ref{PDUsession}. This UE has one PDU session (session-1), which is made up of two DRBs (DRB-1 and DRB-2) on the radio side of gNB and a single tunnel on the NG-U side of gNB. PDU session-1 consists of 3 QoS flows with flow-1 and flow-2 mapped to DRB-1 and flow-3 mapped to DRB-2 on the radio side and using tunnel-1 on the NG-U side. To establish this PDU session, Open5G configures the logical ports in gNB, as indicated in Figure \ref{mapping}. The radio link between UE and d-gNB can be treated as an OF physical port, and each bearer to the UE can be treated as an OF logical port. \par
The Cell Radio Network Temporary Identifier (C-RNTI) of UE available at the base station, along with the bearer-id, can be used to identify the logical port on the radio side of d-gNB. The GTP tunnels corresponding to individual session paths for a UE towards the 5GC are used to identify the logical ports on the NG-U side of d-gNB. The IP address of d-gNB along with the UDP port number and the GTP tunnel-id can be treated as the logical port on the NG-U side of the d-gNB. Open5G Configuration Point, which is a part of SRC configures these session paths using principles similar to OF-Config. However, as mentioned before, OF and OF-Config do not support radio interface and GTP protocols today, and hence, modifications would be required to support them. \par
\begin{figure}
	\centering
	\includegraphics[width=8.6cm]{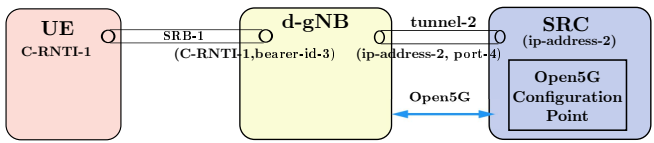}
	\caption{Open5G ports configuration for SRBs.}
	\label{SRBmapping}
\end{figure}
Consider a Signaling Radio Bearer (SRB) setup between UE and SRC. Similar to the data session setup, an SRB can be setup using Open5G. To establish SRB-1 for UE with C-RNTI-1, Open5G configures two logical ports in d-gNB, as illustrated in Figure \ref{SRBmapping}. The radio link between UE and d-gNB with a bearer can be treated as one logical port. The bearer-id, together with C-RNTI-1, can be used to identify the logical port on the radio side of d-gNB. The link from d-gNB to SRC can be treated as another logical port which forms a tunnel to SRC. The IP address of SRC, along with the tunnel-id or port number, identifies the logical port from d-gNB to SRC. An SRB-0 is established in the beginning as a common logical port on RAN data-plane node (d-gNB/d-eNB) for all common channel messages to be exchanged with UEs. A corresponding tunnel is also established between the SRC and d-gNB/d-eNB to carry these UE-specific common channel messages. \par
Analogous to OF, Open5G uses a flow table with (Flow-id, action) tuples to map the flows from ingress to egress logical ports and vice versa. In the uplink, it maps DRBs (logical ports) on the radio side of d-gNB to tunnels (logical ports) on the NG-U side of d-gNB. In the downlink, it maps QoS flows from the NG-U side of d-gNB to their matching DRBs on the radio side of d-gNB. 
For the scenario discussed above, the corresponding flow table is given in Table \ref{flowTable}. The bearer-id represents the ingress port (DRB) where the packets belonging to the flow enter d-gNB on the radio side. For instance, in the case of uplink, it consists of C-RNTI and bearer-id. The action consists of the 'output' action, where the flow gets routed to a specific output port. The output port is a logical port on the NG-U side of d-gNB identified by the IP address of d-gNB along with the UDP port number and the GTP tunnel-id. \par
\begin{table}
\caption{Open5G flow table at d-gNB.}
\begin{center}
\scalebox{0.9}{
	\begin{tabular}{|c|c|}
		\hline
		\textbf{Flow-id } & \textbf{Action} \\
		\hline
		DRB-1\{LP4\} (C-RNTI-1,bearer-id-1) & Output: \{LP1\} port-1,tunnel-id-1\\
		\hline
		DRB-2\{LP5\} (C-RNTI-1,bearer-id-2) & Output: \{LP1\} port-1,tunnel-id-1\\
		\hline
		IP1,TCP,port-43 (Flow-1) & Output: DRB-1\{LP4\} (C-RNTI-1,bearer-id-1)\\
		\hline	
		IP1,TCP,port-23 (Flow-2) & Output: DRB-1\{LP4\} (C-RNTI-1,bearer-id-1)\\
		\hline			
		IP2,TCP,port-34 (Flow-3) & Output: DRB-2\{LP5\} (C-RNTI-1,bearer-id-2)\\
		\hline		
		SRB-1\{LP3\} (C-RNTI-1,bearer-id-3) & Output: \{LP2\} port-4, tunnel-id-2\\
		\hline
		\{LP2\} port-4, tunnel-id-2 & Output: SRB-1\{LP3\} (C-RNTI-1,bearer-id-3)\\
		\hline				
	\end{tabular}
}
\end{center}
\label{flowTable}
\end{table}
In the downlink, the flows are mapped to the corresponding DRB (logical port) identified by C-RNTI of UE and bearer-id. For instance, in downlink Flow-1 is to be output on DRB-1 with C-RNTI-1 and bearer-id-1. In this way, the flow table maps the bearers from the radio side of d-gNB to the tunnel on the NG-U side of d-gNB and vice versa. The flow table also maps the SRBs from the radio side to the SRC side of d-gNB and vice versa (entries 6 and 7 in Table \ref{flowTable}). \par
Thus, in Open5G, OF-Config concept is used to configure the ports, and OF concept is used to route flows through d-gNB based on the port configuration and mapping. The radio bearer (logical port) configuration requires LTE SDAP, PDCP and RLC layer configuration parameters to be supplied by the SRC to d-gNB, while radio link configuration requires LTE MAC and PHY layer parameters to be sent by SRC to d-gNB. Similarly, for the logical port on the NG-U side of d-gNB, the GTP tunnel configuration shall be provided by SRC to d-gNB. 
\section{Working of SMRAN using Open5G}
\label{sec4}
In the previous section, we have seen how the data paths are configured and established using Open5G. In this section, we explain the working of Open5G using certain usecases. 
\subsection{Communication paths for UE over d-gNB}
\subsubsection{Communication path for data}
\begin{figure}[!htb]
	\centering
	\includegraphics[width=6cm,height=5.5cm]{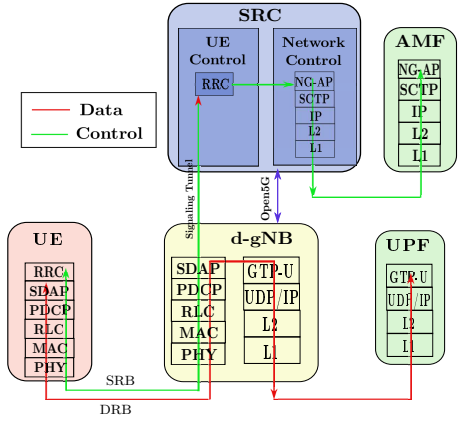}
	\caption{Control and Data signal path.}
	\label{signaling}
\end{figure}
In this subsection, we describe the path followed by the data from a UE to 5GC and vice versa. Figure \ref{signaling} depicts the path followed by data between UE and 5GC in red. In the uplink, data packets belonging to say, bearer-id-1 generated by higher layers at UE (with C-RNTI-1) go through PDCP, RLC, MAC, and PHY layers onto the air interface. The logical port at the radio interface of d-gNB receives these packets and processes them using PHY, MAC, RLC, PDCP, SDAP layers. The flow table at d-gNB (Table \ref{flowTable}) maps these packets from bearer-id-1 to the logical port (LP1) with port-1 and tunnel-id-1. At this port, the packets undergo processing through GTP-U, UDP/IP, L2, and L1 layers. This tunneling sends the packets to UPF. In the downlink, the exact reverse procedure is followed.
\subsubsection{Communication path for signaling}
Figure \ref{signaling} depicts the path followed by control signaling between UE and 5GC in green. In the uplink, signaling packets from higher layers arrive at the RRC layer of UE. These packets are processed using PDCP, RLC, MAC, PHY layers, and transmitted over the radio interface. These are then received at the radio interface of d-gNB. The logical port at the radio side of d-gNB processes the packets on its radio stack (PHY, MAC, RLC, PDCP, SDAP). Using the flow table entry for SRB-1, these packets are then forwarded through the signaling tunnel (GRE or it can be other tunneling protocol also) created for SRB-1 to SRC. SRC then internally processes these packets and sends them via NG-AP, SCTP, IP, L2, and L1 to AMF through the NG-C interface.
\subsection{Procedure for UE initial access in SMRAN}
\begin{figure}
	\centering
	\includegraphics[width=9cm]{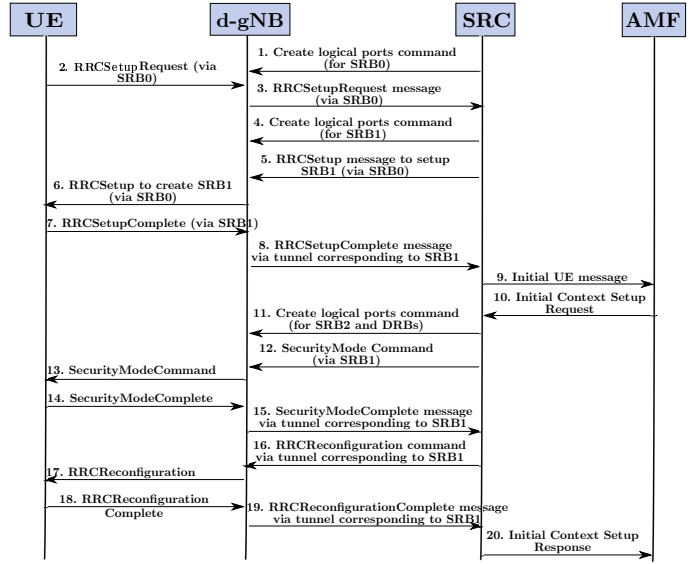}
	\caption{Call flow procedure for UE initial access.}
	\label{callFlow}
\end{figure}
In this section, we describe the procedure for UE initial access in SMRAN, as illustrated in Figure \ref{callFlow}. Each message is explained below.\par
1) Initially, on system startup, Open5G configuration point in SRC issues the command to create and configure ports for SRB0 at d-gNB and maps them in the flow table so that path (SRB0) is created between UE and SRC. This command uses \textit{ofp\_port\_mod} message of OF \cite{OF} to configure the ports and \textit{OFPT\_FLOW\_MOD} message of OF to create an entry in the flow table to map these logical ports. SRB0 is common for all UEs and is created on system startup only once. 2) When a UE comes in the coverage area of d-gNB, it sends an RRCSetupRequest to d-gNB via SRB0. 3) d-gNB sends this RRCrequest to SRC through the tunnel corresponding to SRB0, between the d-gNB and SRC. 4) SRC, then decides if UE has to be admitted to the network. If UE is admitted, SRC sends Create logical ports command with details of SRB1 to be created to d-gNB. \par
5) SRC sends the RRCSetup message to d-gNB via the tunnel corresponding to SRB0 to setup SRB1. 6) d-gNB sends RRCSetup message to UE for creating SRB1 through SRB0. 7) SRB1 is now set up. UE sends RRCSetupComplete message to d-gNB over SRB1, which also contains the NAS message for AMF. 8) The d-gNB sends the RRCSetupComplete message to SRC via the tunnel corresponding to SRB1. 9) SRC sends the Initial UE message to the AMF. 10) AMF replies with initial context setup request along with details of PDU sessions to be established and security information for UE. 11) SRC sends Create logical ports command with details of SRB2 and DRBs to be established to the d-gNB. 
12) SRC sends the SecurityModeCommand to d-gNB through the tunnel corresponding to SRB1. \par
13) d-gNB sends the SecurityModeCommand message to the UE through SRB1. 14) UE responds with the SecurityModeComplete message to d-gNB through SRB1. 15) The d-gNB sends the SecurityModeComplete message to the SRC via the tunnel established for SRB1. 16) The SRC generates the RRCReconfiguration message and sends it to the d-gNB via the SRB1 tunnel. 17) The d-gNB sends RRCReconfiguration message to the UE via SRB1. 18) The UE reconfigures its RRC and sends RRCReconfigurationComplete message to the d-gNB. 19) The d-gNB sends the RRCReconfigurationComplete message to the SRC via the tunnel established for SRB1. 20) SRC sends Initial Context setup Response to AMF containing details of the PDU session resources setup.\par
\section{Advantages of Open5G and SMRAN}
\label{sec5}
\begin{itemize}
	\item SMRAN is in alignment with the current 5G architecture with the control-plane of all RAN nodes hosted at the SRC. This motivates the use of Open5G as a singular protocol for communication between SRC and data-plane entities belonging to different RATs.
	\item SMRAN unifies multi-RAT RAN and obviates the need for RAT-specific functions like N3IWF in RAN proposed to be used for WLAN integration with 5GC.
	\item SMRAN enables integration of LTE, 5G, and WLAN with the 5GC in a unified manner.
	\item SMRAN architecture ensures that UE is not impacted, and its interfaces with the network remain the same.
	\item OpenFlow protocol exists in SDN-based wired networks. With the introduction of Open5G, OpenFlow can be incorporated in the RAN of wireless networks as well. This work can be extended to support OpenFlow in the core network. The usage of a single protocol in the network can then simplify the architecture and control management procedures. 
	\item With Open5G, we have a unified interface for control and management of RAN with multi-RAT technologies such as WLAN, LTE, 5G.
	\item With the introduction of Open5G and SMRAN, one interface (E1) has been eliminated. 
	\item The centralized control at SRC enables it to command the RAN nodes using Open5G. Hence, the response messages from RAN nodes are not required. This enables a reduction in the messages to be exchanged between SRC and RAN nodes.	
	\item The F1 interface between gNB-CU and gNB-DU in 5G has been changed to Open5G in SMRAN. The main functions of the F1 interface are to carry UE-specific signaling messages and to configure gNB-DU. However, Open5G commands are used to only configure the RAN nodes (e.g., d-gNBs) to establish data paths between the UE and network nodes. It also means that Open5G is responsible for setting up the paths to carry both UE specific data as well as signaling messages, treating both, UE signaling and data, in a similar fashion. In the SMRAN architecture, the UE-specific signaling messages are sent from the SRC to UE via the d-gNB in the same manner as the UE specific data from the UPF. This simplifies the flow of different types of traffic within the network. One of the objectives of SDN is to program the network entities as per the requirements. With a centralized controller to manage and control the RAN nodes, we separate the network configuration commands from UE-specific messages, and thus, achieve the objectives of SDN. This also brings modularity to the SMRAN architecture.
\end{itemize}
\section{Conclusions and Future Work}
\label{conc}
We propose an SDN-based Multi-RAT RAN (SMRAN) architecture, which is in alignment with the 5G architecture. In SMRAN, the control-plane functions of all RAN nodes are aggregated at a centralized controller. A global view of the RAN at the controller enables the design of efficient association and load balancing algorithms. SMRAN unifies multi-RAT RAN and does away with the RAT-specific functions such as N3IWF in RAN. We propose the Open5G protocol for communication between the controller and RAN nodes. Open5G uses OpenFlow and OF-Config, which are the commonly used protocols in SDN-based wired networks. With Open5G, the same procedures can be used to control and manage RAN nodes belonging to different RATs. We detail the Open5G protocol and also elaborate on the data and control paths followed in Open5G. We also describe the call flow procedure to be followed in SMRAN for initial access of a UE. The usage of a protocol similar to Open5G in the core network can be taken up as a part of future work. For future work, we would also like to evaluate the performance of Open5G in wireless networks.
%
\bibliographystyle{ieeetr}
\bibliography{IEEEabrv,MOF_References}

\begin{thebibliography}{10}

\bibitem{Pradnya}
P.~K. Taksande, P.~Jha, and A.~Karandikar, ``Dual connectivity support in {5G}
  networks: An {SDN} based approach,'' in {\em IEEE Wireless Communications and
  Networking Conference}, pp.~1--6, 2019.

\bibitem{38470}
3GPP, ``{NGRAN; F1 general aspects and principles (Release 15)},'' TS 38.470,
  {3rd Generation Partnership Project}, Jul 2019.
\newblock Version 15.6.0.

\bibitem{38460}
3GPP, ``{NGRAN; E1 general aspects and principles (Release 15)},'' TS 38.460,
  {3rd Generation Partnership Project}, Jul 2019.
\newblock Version 15.4.0.

\bibitem{38401}
3GPP, ``{NGRAN; Architecture description},'' TS 38.401, {3rd Generation
  Partnership Project}, Jul 2019.
\newblock Version 15.6.0.

\bibitem{mobileflow}
K.~Pentikousis, Y.~Wang, and W.~Hu, ``Mobileflow: Toward software-defined
  mobile networks,'' {\em IEEE Communications magazine}, vol.~51, no.~7,
  pp.~44--53, 2013.

\bibitem{softair}
I.~F. Akyildiz, P.~Wang, and S.-C. Lin, ``{SoftAir:} a software defined
  networking architecture for {5G} wireless systems,'' {\em Computer Networks},
  vol.~85, pp.~1--18, 2015.

\bibitem{oran}
O.~R. Alliance, ``{O-RAN:} towards an open and smart {RAN},'' {\em White Paper,
  October}, 2018.

\bibitem{5gempower}
E.~Coronado, S.~N. Khan, and R.~Riggio, ``{5G-EmPOWER:} a software-defined
  networking platform for {5G} radio access networks,'' {\em IEEE Transactions
  on Network and Service Management}, 2019.

\bibitem{flexran}
X.~Foukas, N.~Nikaein, M.~M. Kassem, M.~K. Marina, and K.~Kontovasilis,
  ``{FlexRAN: A} flexible and programmable platform for software-defined radio
  access networks,'' in {\em {ACM Conference on emerging Networking Experiments
  and Technologies (CoNEXT)}}, pp.~427--441, 2016.

\bibitem{akshatha}
N.~M. Akshatha, P.~Jha, and A.~Karandikar, ``A centralized {SDN} architecture
  for the {5G} cellular network,'' in {\em {IEEE 5G World Forum (5GWF)}},
  pp.~147--152, 2018.

\bibitem{hierarchical}
G.~Yu, R.~Liu, Q.~Chen, and Z.~Tang, ``A hierarchical {SDN} architecture for
  ultra-dense millimeter-wave cellular networks,'' {\em IEEE Communications
  Magazine}, vol.~56, no.~6, pp.~79--85, 2018.

\bibitem{akshatha2}
N.~M. Akshatha, A.~Roy, P.~Jha, and A.~Karandikar, ``Control and management of
  multiple {RATs} in wireless networks: An {SDN} approach,'' in {\em IEEE 5G
  World Forum (5GWF)}, pp.~596--601, 2019.

\bibitem{OF}
{Open Networking Foundation}, ``Openflow switch specification v1.5.1,'' 2015.

\bibitem{OnfSdn}
{Open Networking Foundation}, ``Software-defined networking: The new norm for
  networks,'' {\em White Paper}, vol.~2, pp.~2--6, 2012.

\bibitem{OF-C}
{Open Networking Foundation}, ``Openflow management and configuration protocol
  v1.2,'' 2014.

\bibitem{mimoOF}
S.~Kumar, D.~Cifuentes, S.~Gollakota, and D.~Katabi, ``Bringing cross-layer
  {MIMO} to today's wireless {LANs},'' in {\em ACM SIGCOMM Computer
  Communication Review}, vol.~43, pp.~387--398, 2013.

\bibitem{openroads}
K.-K. Yap, M.~Kobayashi, R.~Sherwood, T.-Y. Huang, M.~Chan, N.~Handigol, and
  N.~McKeown, ``{OpenRoads:} empowering research in mobile networks,'' {\em ACM
  SIGCOMM Computer Communication Review}, vol.~40, no.~1, pp.~125--126, 2010.

\bibitem{38300}
3GPP, ``{NR; Overall description; Stage-2},'' TS 38.300, {3rd Generation
  Partnership Project}, Jan 2018.
\newblock Version 15.0.0.

\end{thebibliography}

\end{document}